\documentstyle[pramana,floats]{ias}
\begin{document}
\mark{{Absence of M2..$^{35}Cl$...}{Ritesh Kshetri et al.}}
\title{
Absence of M2 Retardation in $^{35}Cl$: Evidence for Stronger Isospin-Mixing
 Effects in A=35 Mirror Nuclei}
\author{Ritesh Kshetri, Indrani Ray, P. Banerjee, R. Raut, A. Goswami, J. M. 
Chatterjee, S. Chattopadhyay, U. Datta Pramanik, A. Mukherjee, C.C. Dey, 
 S. Bhattacharya, B. Dasmahapatra and M. Saha Sarkar}
\address{Saha Institute of Nuclear Physics, 1/AF, Bidhannagar, Kolkata - 700064}
\author{S. Sarkar}
\address{Department of Physics, The University of Burdwan, Golapbag, 
Burdwan -713104}
\author{S. Bhowal}
\address{Surendranath Evening College, Kolkata - 700016}
\author{G.  Ganguly}
\address{Department of Physics, The University of Calcutta, Kolkata - 700009}
\author{K. S. Golda, R. Kumar, R. P. Singh, S. Muralithar, P. V. Madhusudhana 
Rao, N.  Madhavan,   J. J. Das,  S. Nath,  P. Sugathan,  A. Jhingan, R. K. 
Bhowmik}
\address{Nuclear Science Centre, Aruna Asaf Ali Marg, New Delhi -110067}
\author{P. Datta}
\address{Anandamohan College, Kolkata- 700009}
\author{P.K. Joshi,  H.C. Jain} 
\address{Tata Institute of Fundamental Research,  Mumbai-400005}
\keywords{A=40, gamma spectroscopy, lifetime}
\pacs{21.10.Tg,27.30.+t}
\abstract{
 The lifetime of  the 3163 keV, 7/2$^-$ isomeric  state in $^{35}Cl$  
 that decays by a stretched M2  transition to the $3/2^+$ ground state, 
 has been re-measured using the Doppler Shift Attenuation Method, by gating  on the
 1185  keV  transition which  directly  feeds this  state.  This eliminates  the
 uncertainties in  the measurement  arising from  the direct  feedings from  the
 continuum. A mean life of 0.6$^{+0.5}_{-0.2}$ ps has been obtained from the present work.
This is  considerably smaller than the adopted value 45.3(6) ps. Implication  of  this  major
 reduction in the lifetime has been pointed out.}
 
\maketitle
\section{Introduction}
Low-lying  isomeric states  that decay  by stretched  M2 transitions  have been
observed in  nuclei near  the doubly  closed $^{40}Ca$.  During late  sixties to
early eighties several measurements of  lifetimes of these isomeric states  have
been performed  by populating  them in  fusion evaporation  reactions with heavy
ions or  by using  protons. Electronic  timing techniques  and different Doppler
shift methods were used for the measurements.  These M2 transitions are  usually
explained using the shell model.  Their interpretation is connected with  the de
-excitation of a single nucleon  from the  1$f_{7/2}$ to  the 1$d_{3/2}$  orbit. The
single-nucleon transfer reactions have revealed large 1$f_{7/2}$ single particle
components (spectroscopic factor S$\simeq$ 0.54 on  the average) in the wavefunction  of
the  lowest 7/2$^-$  state or  their analogues.  The information  on the  lowest
3/2$^+$ state in  odd K, Ca,  Sc, etc. indicated  sizable d$_{3/2}$ single  hole
component.   But in  spite of  the dominant  particle hole  character of  these
levels, a systematic hindrance of the  M2 transitions by a factor of  5-200 with
respect to 1$f_{7/2}  \rightarrow 1d_{3/2}$ single  particle estimates has  been
observed.  Several  attempts  have  been  made  to  account  for  this anomalous
retardation effect. For the relatively heavier  nuclei with A=39, 41 and the  Sc
isotopes, isospin effects and/or admixtures  of core excited states can  explain
this retardation. Nuclei, which are three or more particles away from the  core,
need strong coupling and prolate deformation to account for
the observed retardation.  

Here we concentrate   on the re-measurement of the lifetime of 7/2$^-$ isomeric state
in the $^{35}Cl$ nucleus  \cite{War:0} using the Doppler Shift Attenuation Method (DSAM).
A mean life of 0.6$^{+0.5}_{-0.2}$ ps has been obtained from the present work.
This is  considerably smaller than the adopted value 45.3(6) ps.
This  present value indicates the absence of  M2 retardation  in this
nucleus.  The new result has far reaching implications which will be discussed.

\section{Experimental Details}

The present work is based on the results from two experiments done with the array 
of eight  Clover detectors (INGA setup)  at TIFR, Mumbai (Expt. 1) and  NSC, New
Delhi (Expt. 2). A 50 $\mu$g/cm$^2$  $^{12}C$ target, backed  by  $\simeq$  10.5
mg/cm$^2$ gold was bombarded by 70   and 88 MeV $^{28}Si$ beam in the two experiments, 
respectively.  The velocities  of the recoils were 5-6\% of velocity of light ($c$). The main  
interest was to study  the higher  spin  states  in the nuclei $^{38}Ar$,  $^{35}Cl$ and
$^{37}Ar$ and  other weaker channels  like   $^{35}Ar$,  $^{38}K$  where heavy  
ion  data  are  scanty. DSAM
was  used to  measure the  lifetimes of   several nuclear   states. In Expt. 1
the  detectors were  at 30$^o$, 60$^o$, 65$^o$, 90$^o$, 105$^o$,  120$^o$,
145$^o$ w.r.t the beam, while in Expt. 2, they were at 81$^o$ and 134$^o$ with respect 
to the beam direction.  

Use  of "inverse" reactions in DSA measurements  has  some distinct
advantages. These include large recoil  velocities  and small percentage  spread in  
velocity \cite{War:1}.  For the  $^{12}C(^{28}Si,  \alpha
p)^{35}Cl$  reaction populating  the 3163  keV level  at 88  MeV energy  of  the
$^{28}Si$ beam, the recoil velocity $\beta$ $\simeq$ 5.8\%, the  maximum spread in  
$\beta$ is $\simeq  \pm 7\%$, while  the maximum half angle of the cone  of recoiling 
$^{35}Cl$ nuclei is $\simeq  13^o$, compared to the value  2.5\%, $\pm 17\%$, and 
 31$^o$, respectively, for the corresponding  "forward" reaction at the same centre
of mass energy. Due to the larger velocity, the  
fractional energy loss of the recoils in  the target material will be comparatively  
smaller, thereby reducing the uncertainty due  to finite  target  thickness. 
Besides, there is more 
freedom in choosing the backing material.
Also  for  higher   recoil   velocity,  the   main contribution to the stopping is  the 
electronic stopping process, which  is well understood. 

However, due to the larger recoil velocity, the lineshape extends over a  large number 
of channels (e.g., for 3163
keV gamma  ray, with  $\beta_{max} =0.0575$  and $\theta$  = 134$^o$,  the endpoint
corresponds to  a shift  of 126  keV). The  background under  this shape  can be
sizable, even when it does not  show any noticeable structure. Very often  there
are structures (contribution from other genuine or spurious peaks) which interfere 
with the lineshape even  in the gated spectrum. 

\section{Data Analysis and Results}
In the present work we have made a  preliminary estimate of the   lifetime of
the 7/2$^-$ isomeric state in $^{35}Cl$ nucleus using the centroid shift analysis 
of the DSAM data.  A strong 3163 keV M2 transition connects this state  to
the  3/2$^+$  ground  state.  The  lifetime of   the  3163  keV  state  has been
estimated by gating on the 1185  keV transition which directly  feeds the  state
of   interest   \cite{War:0}.  This   eliminates   the   uncertainties   in  the
measurement  arising   from  the   direct  feedings  from  the   continuum.

\subsection{Energy loss of recoil}
The energy loss of the $^{35}Cl$  nuclei through $^{12}C$ target was very small for  both
energies (70 and 88 MeV) and hence was neglected. The energy loss  of $^{35}Cl$
in gold has been  simulated using the code SRIM-2003  \cite{Zie:1}.
The energy loss of the recoils is parametrised by using the relation \cite{War:2}, 
\begin{equation}
-M(dv_z/dt)=K_n (v_z/v_o)^{-1} + K_e (v_z/v_o) - K_3 (v_z/v_o)^{3}
\end{equation}
where M is the mass  of the moving ion, $v_o$=  $c$/137. $z$  direction is along  the
path of the recoil. The values  of the parameters $K_n=0.1$,  $K_e=1.1$
and $K_3=0.006$ have been  obtained by fitting the  simulated data by the  above
relation. These parameters were used in calculating the theoretical values of the
attenuation  coefficient F($\tau$) \cite{War:2}, (to  be discussed  in the  next section) as a
function of mean lifetime ($\tau$) of the  level emitting that particular gamma ray. 

\subsection{Determination of experimental F$(\tau)$}

We shall assume that the recoils are all emitted along the z  axis
and they are monoenergetic. 
The average energy  \=E$_\gamma$ of the gamma radiation emitted from an
ensemble of nuclei produced at t=0 with initial velocity $\beta$(0) ($= v(0)/c$)
and   moving  thereafter   with  velocity  $\beta$(t)   can  be  expressed   for
$\beta(t)\ll 1$ as \cite{War:2}, 

\begin{equation}
\overline{E_\gamma} = E_{\gamma o} [1+F(\tau) \beta(0)cos\theta]
\end{equation}

$E_{\gamma o}$ is the $\gamma$-energy  emitted by the recoil at  rest. F$(\tau)$
is the attenuation coefficient  which lies between 0  and 1. The lifetime  of the
level emitting that particular gamma ray can be determined if F$(\tau)$  differs
measurably from 1 and 0. It can be obtained from the observed energy shift  with
the detector angle of the gamma, using the relation
 
\begin{equation}
F(\tau)={\Delta E_\gamma \over E_{\gamma o} \beta (0) (cos\theta_1-cos\theta_2)}
\end{equation}

The F$(\tau)$  value thereby obtained has been corrected for the effective lifetime of the 4348 
keV level which decays by the 1185 keV $\gamma$ ray to the level of interest, since this transition
has been used as the gating transition. The centroids were
determined after consideration of suitable background using the analysis program
INGASORT \cite{Rkb:1}. The experimental results are summarised in Table 1.

\begin{table}[h]
\begin{center}
\caption{Summary of the F$(\tau)$ values and corrected $\tau$(ps) of 3163 keV level.}
\begin{tabular}{llllllll}
Energy &$\beta_{max}$& $\theta_1$&$\theta_2$&$E_\gamma$&$\Delta E$&F$(\tau)$& Corrected $\tau$  \\ 
(MeV)  &             &           &          & (keV)& (keV) &   & (ps) \\ \hline

88 &0.0575& 81$^o$& 134$^o$&3163&14&0.09& 0.5 \\
70 &0.0513& 90$^o$& 105$^o$&3163&3&0.07& 0.6  \\
   &      & 90$^o$& 120$^o$&3163&7&0.09&  \\
\end{tabular}
\end{center}
\end{table}

\subsection{Limitation in  previous measurements  and estimation  of error in
present measurement}

The large lifetime of 45.3 ps reported \cite{nnd:1} previously may possibly be 
attributed to one or more of the following limitations:

\begin{itemize}
\item{} Use of evaporated  targets with layered structures or  backing
materials whose stopping powers were poorly known.  
\item{} Population of the level of interest in forward reactions producing low recoil 
velocities (in most of the cases $\beta$ $\leq$ 1\%). 
\item{} Measurements  in  singles mode are difficult in A$\simeq$ 40 due to 
interference from overlapping gamma rays.
\item{} The single detector resolution and efficiency at high energies were poor.
\end{itemize}

The present measurements were done using an array which consisted of eight clover detectors.
These detectors have good resolution and efficiency at high energies.

\begin{itemize}
\item{}  Although the present measurement also had some serious limitations, like 
inadequate backing thickness for Expt. 2 (resulting in an escape of 
$\simeq$ 5\% of recoils) and the difficulty 
in setting a clean gate on the 1185 keV lineshape, the associated uncertainties have been 
estimated and included in the errors in the lifetime result.
\item{} Our energy calibration was 1 keV/channel, and therefore an error  of $\pm$1
channel in the determination of the centroid introduced large errors in F($\tau$)
especially for comparatively lower  energies. As  the gamma  energies associated
with  this mass region  are usually of high  energies ($> 1$MeV, even  up to 5-6
MeV), we were bound to compress the spectrum.
\end{itemize}

A  lineshape analysis of the data is in progress which is expected to yield a more accurate result.
However, the present work establishes that the lifetime of the 3163 keV state in $^{35}Cl$ is 
considerably less than the value reported in the literature.

\section{Theoretical analysis and implications}

The adopted value \cite{nnd:1} of  the  mean  lifetime of  3163 keV  level in
$^{35}Cl$  ($7/2^-$)  is  45.3  (6)  ps.  The  experimental  reduced  transition
probabilities of the M2 transitions  has  been  calculated using the relation given in
Ref. \cite{Kei:1}. Taking branching ratio of the 3163  keV transition to  be 90\%,
mixing  ratio $\delta$ (E3/M2) = -0.16 \cite{nnd:1},  and the  gamma energy  as 3163  keV, the 
calculated B(M2)$_{old-exp}$ value comes out to  be 4.72 $\mu_o^2fm^2$. It is 18 times retarded
compared  to the single particle estimate of B(M2) (using relation mentioned in 
Ref. \cite{Boh:1}) for a stretched  M2 transition. The present measured  value  of the  
lifetime  of 3163  keV  level of 0.6$^{+0.5}_{-0.2}$  ps leads to B(M2)$_{new-expt.}$ $\simeq$  
200 $\mu_o^2fm^2$, which is nearly 2.5 times the single particle estimate.

It  has been  discussed in Ref. \cite{Ekm:1} that  isospin symmetry breaking  effects can  be
studied in pairs of mirror  nuclei, in which the number of protons and  neutrons
are  interchanged.  They lead  to shifts  between the  excitation  energies of a
mirror pair, the  so­ called mirror  energy  differences (MED).  However, it has
long  been expected and recently been  shown that a  small part of  the nucleon­
-nucleon  interaction  adds  to  the  Coulomb  force  in  violating  the isospin
symmetry \cite{Ekm:1}.  Ekman {\it et al.} made a comparison  of  $^{35}Ar$  (T$_z$=-1/2)  with the
T$_z$=1/2 mirror nucleus $^{35}Cl$. It  reveals  two remarkable features: (i) A  very
large MED value for the 13/2$^-$ states,  and (ii) a dramatic difference in  decay
patterns of the 7/2$^-$ states.  The explanation for the dramatic  difference in
decay  patterns  of the  7/2$^-$   states in  the  A =35  mirror  pair has  been
explained to be due to a cancellation  of the E1 matrix elements due to  isospin
mixing. Using the adopted mean  lifetime $\tau$=45.3(6) ps of the  7/2$^-$ state
and B(E1;7/2$^- \rightarrow  5/2^+$)=2 $\times 10^{-8}$  W.u. in $^{35}Cl$,  and
assuming identical  B(M2)'s in  both members  of the  mirror system,  it follows
that B(E1;7/2$^- \rightarrow 5/2^+$)=3 $\times 10^{-5}$ W.u for $^{35}Ar$.  This
means  that  the contributions  diagonal  and non-diagonal  in isospin, $T$ are  equal  in
magnitude and about 1.5$\times 10^{-5}$  W.u \cite{Ekm:1}. For present value of  B(M2)$_{exp}$
$\simeq$  200 $\mu_o^2fm^2$,  $\tau$ comes  out to  be $\simeq$ 0.15 ps for $^{35}Ar$  and
B(E1;7/2$^-   \rightarrow   5/2^+$) $\simeq$ 6$\times10^{-7}$  W.u   for   $^{35}Cl$  and
B(E1;7/2$^- \rightarrow  5/2^+$) $\simeq$ 1.5$\times 10^{-3}$  W.u for  $^{35}$Ar. The present 
results indicate the need for a stronger isospin mixing. In fact, a more detailed experimental and theoretical 
investigation is needed to resolve this important issue.

\end{document}